\documentstyle[twoside,fleqn,npb,epsfig]{article}
\newcommand{\Li}[2]{{\mbox{Li}}_{#1}\left(#2\right)}
\newcommand{\snp}[2]{{\mbox{S}}_{#1}\left(#2\right)}

\newcommand{\nigelboxd}{I_{Irr}^d}
\newcommand{\svboxad}{I_1^d}
\newcommand{\svboxbd}{I_2^d}
\newcommand{\diagboxd}{I_{Diag}^d}
\newcommand{\bubboxd}{I_{Bbox}^d}
\newcommand{\glassd}{I_{Spec}^d}
\newcommand{\trid}{I_{Btri}^d}
\newcommand{\sunsetd}{I_{Sunset}^d}

\newcommand{\nigel}{I_{Irr}}
\newcommand{\smirnov}{I_1}
\newcommand{\veretin}{I_2}
\newcommand{\diagb}{I_{Diag}}
\newcommand{\blobb}{I_{Bbox}}
\newcommand{\spec}{I_{Spec}}
\newcommand{\tr}{I_{Btri}}
\newcommand{\suns}{I_{Sunset}}

\def\SUNC{
\setlength{\unitlength}{1cm}
\mbox{\parbox{2.1cm}{\hspace{0.25cm}
\begin{picture}(0.1,1)
\thicklines
\put(0.1,0.5){\line(1,0){1.8}}
\put(1,0.5){\circle{0.9}}
\end{picture}
}}
\hfill}

\def\GLASS{
\setlength{\unitlength}{1cm}
\mbox{\parbox{3.1cm}{\hspace{0.25cm}
\begin{picture}(3.1,1)
\thicklines
\put(0.15,0.5){\line(1,0){0.5}}
\put(2.4,0.5){\line(1,0){0.5}}
\put(1.1,0.5){\circle{0.9}}
\put(1.95,0.5){\circle{0.9}}
\end{picture}
}}
\hfill}

\def\triA{
\setlength{\unitlength}{1cm}
\mbox{\parbox{2.4cm}{\hspace{0.25cm}
\begin{picture}(2,1.4)
\thicklines
\put(0.05,0.7){\line(1,0){0.5}}
\put(1,1.14){\line(0,-1){0.85}}
\put(1,1.13){\line(1,0){0.8}}
\put(1,0.28){\line(1,0){0.8}}
\put(1,0.7){\circle{0.9}}
\end{picture}
}}
\hfill}

\def\Abox{
\setlength{\unitlength}{1cm}
\mbox{\parbox{2.8cm}{\hspace{0.25cm}
\begin{picture}(2.8,1.4)
\thicklines
\put(0.2,0.28){\line(1,0){2.4}}
\put(0.2,1.13){\line(1,0){2.4}}
\put(0.7,0.28){\line(0,1){0.85}}
\put(1.7,0.7){\circle{0.9}}
\end{picture}
}} 
\hfill}

\def\Cbox{
\setlength{\unitlength}{1cm}
\mbox{\parbox{2.5cm}{\hspace{0.25cm}
\begin{picture}(2.5,1.4)
\thicklines
\put(0.25,0.2){\line(1,0){2}}
\put(0.75,0.2){\line(1,1){1}}
\put(0.25,1.2){\line(1,0){2}}
\put(0.75,0.2){\line(0,1){1}}
\put(1.75,0.2){\line(0,1){1}}
\end{picture}
}} 
\hfill}

\title{The on-shell massless planar double box diagram with an
       irreducible numerator}

\author{C.~Anastasiou,\address{
        Department of Physics,
        University of Durham, Durham DH1 3LE, England}\thanks{
        \tt Ch.Anastasiou@durham.ac.uk}
        J.B.~Tausk\address{
        Fakult{\"a}t f{\"u}r Physik,
        Albert-Ludwigs-Universit{\"a}t Freiburg,
        D-79104 Freiburg, Germany}\thanks{
        \tt tausk@physik.uni-freiburg.de}
        and
        M.E.~Tejeda-Yeomans{\hbox{$^{\rm a}$}}\thanks{
        \tt M.E.Tejeda-Yeomans@durham.ac.uk}
}

\begin{document}

\begin{abstract}
Using a Mellin-Barnes representation, we compute the on-shell massless
planar double box Feynman diagram with an irreducible scalar product of
loop momenta in the numerator. This diagram is needed in calculations
of two loop corrections to scattering processes of massless particles,
together with the double box without numerator calculated previously
by Smirnov. We verify the poles in $\epsilon$ of our result by means of
a system of differential equations relating the two diagrams, which we
present in an explicit form. We verify the finite part with an independent
numerical check.
\end{abstract}

\maketitle

\thispagestyle{myheadings}
\markright{DTP/00/32, Freiburg-THEP 00/9, hep-ph/0005328}


\section{Introduction}

\begin{figure}[htb]
\centerline{\epsfig{file=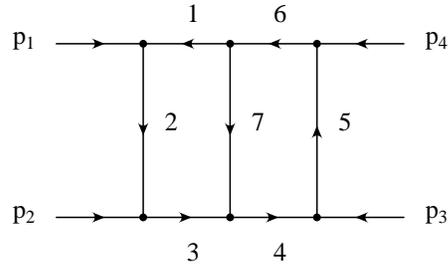,width=6cm}}
\caption{The planar double box diagram}
\label{fig:pbox}
\end{figure}

We shall consider the following class of Feynman integrals
\begin{eqnarray}
\label{eq:Idef}
\lefteqn{
I(\nu_1,\nu_2,\nu_3,\nu_4,\nu_5,\nu_6,\nu_7,\nu_8;d) =
\int \int \mbox{d}^d k \; \mbox{d}^d l \;
}
&& \nonumber \\ &&
\frac{1}{P_1^{\nu_1} P_2^{\nu_2} P_3^{\nu_3} P_4^{\nu_4}
         P_5^{\nu_5} P_6^{\nu_6} P_7^{\nu_7} P_8^{\nu_8}  } ,
\end{eqnarray}
where the propagator denominators $P_i$ are defined by
$P_1 = k^2$,
$P_2 = (k+p_1)^2$,
$P_3 = (k+p_1+p_2)^2$,
$P_4 = (l+p_1+p_2)^2$,
$P_5 = (l+p_1+p_2+p_3)^2$,
$P_6 = l^2$,
$P_7 = (l-k)^2$,
$P_8 = (k+p_1+p_2+p_3)^2$,
$d=4-2\epsilon$ is the space-time dimension, and the external
momenta $p_1$, $p_2$, $p_3$, and $p_4=-p_1-p_2-p_3$ are lightlike.
The integrals~(\ref{eq:Idef}) then depend on two scales: $s=(p_1+p_2)^2$
and $t=(p_2+p_3)^2$. We shall mainly be concerned with cases where $\nu_8$
is either zero or a negative integer. These integrals correspond to the
diagram shown in figure~\ref{fig:pbox}, and are important for
calculations of two-loop scattering amplitudes of massless particles,
for example in QCD \cite{millennium,nigel}.

\begin{figure}[htb]
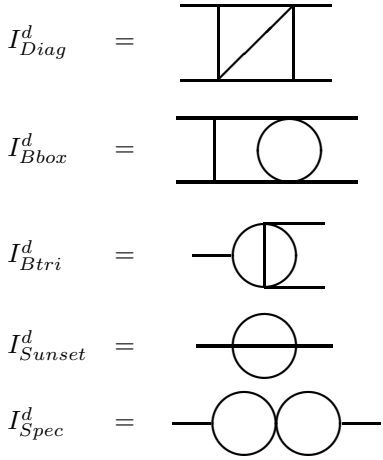

$$
\begin{array}{lcc}
\diagb^d & = & \Cbox  \\
\blobb^d & = & \Abox  \\
\tr^d    & = & \triA  \\
\suns^d  & = & \SUNC  \\
\spec^d  & = & \GLASS
\end{array}
$$
\caption{Master integrals with less than six propagators}
\label{fig:pinchings}
\end{figure}

The scalar double box integral
\begin{equation}
\label{eq:I1def}
\svboxad(s,t) = I(1,1,1,1,1,1,1,0;d) ,
\end{equation}
was calculated analytically by Smirnov \cite{Smirnovdoublebox}
as a Laurent expansion in $\epsilon$ to ${\cal O}(\epsilon^0)$.
Subsequently, Smirnov and Veretin constructed a general algorithm
\cite{SmirnovVeretin99} to deal with tensor double box integrals.
In this algorithm, integrals with arbitrary polynomials of
loop momenta in the numerator are first expressed in terms
of scalar integrals
$I(\nu_1,\nu_2,\nu_3,\nu_4,\nu_5,\nu_6,\nu_7,0;d+2n)$ with increased
integer powers of the propagators in shifted space-time dimensions
\cite{Tarasov}. Then, using integration by parts identities \cite{ibp},
these scalar integrals are systematically reduced to linear
combinations of a set of master integrals, consisting of
$\smirnov$, a second master double box integral,
\begin{equation}
\label{eq:I2def}
\svboxbd(s,t) =  I(1,1,1,1,1,1,2,0;d) ,
\end{equation}
and a small number of more simple master integrals
(see figure~\ref{fig:pinchings}).
Furthermore, they gave an explicit analytical expression for the
second master integral $\veretin$, again as a Laurent expansion in
$\epsilon$ to ${\cal O}(\epsilon^0)$.

It appeared that the algorithm described in ref.~\cite{SmirnovVeretin99}
completely solved the problem of calculating on-shell double box
diagrams. However, as was reported by Glover at this workshop
\cite{nigel}, it often happens that in the reduction of a given
tensor integral, the coefficients in front of the master integrals
$\smirnov$ and $\veretin$ are of order ${\cal O}(1/\epsilon)$. This is
a consequence of the fact that in the reduction of these integrals
it is necessary to reduce the dimension down to $d=4-2\epsilon$
from at least $d=6-2\epsilon$, and in the system of equations for the
dimensional shift, there are factors of $1/(d-6)$ sitting in front of
$\smirnov$ and $\veretin$. Thus, in order to calculate such tensor
integrals to ${\cal O}(\epsilon)$, one would need to know $\smirnov$
and $\veretin$ to one order higher in $\epsilon$ than they are given
in refs.~\cite{Smirnovdoublebox,SmirnovVeretin99}.

A typical example is the following integral with an irreducible
numerator:
\begin{equation}
\label{eq:I3def}
\nigelboxd(s,t) = I(1,1,1,1,1,1,1,-1;d) ,
\end{equation}
whose reduction to master integrals reads \cite{nigel}:
\begin{eqnarray}
\label{eq:relmasters}
\lefteqn{
\nigelboxd(s,t) = -\frac{1}{2}\frac{(3d-14)s}{d-4}\svboxad(s,t)
}
&& \nonumber \\ &&
- \frac{1}{2}\frac{(d-6)st}{(d-4)(d-5)}\svboxbd(s,t)
\nonumber \\ &&
- 3\frac{(s+t)}{(d-5)(d-6)s^{2}t}
\nonumber \\ &&
\times\left[(7d^2-68d+164)s\right.
\nonumber \\ &&
\left.+(3d-14)(3d-16)t\right]\diagboxd(s,t)
\nonumber \\ &&
+24\frac{(d-3)}{(d-6)t}\bubboxd(s,t)
\nonumber \\ &&
-4\frac{(d-3)^2(2d-9)}{(d-4)^2(d-5)s^2}\glassd(s)
\nonumber \\ &&
+\frac{3}{2}\frac{(d-3)(3d-10)}{(d-4)^2(d-5)^2(d-6)s^2t}
\nonumber \\ &&
\times\left[8(d-4)(d-5)^2s \right.
\nonumber \\ &&
\left.+(-11d^3+158d^2 \right.
\left.-754d+1196)t\right]\trid(s)
\nonumber \\ &&
+3\frac{(d-3)(3d-8)(3d-10)}{(d-4)^3(d-5)^2(d-6)s^3t}
\nonumber \\ &&
\times\left[(d-5)(7d^2-68d+164)s \right.
\nonumber \\ &&
\left.+(23d^3-337d^2 \right.
\left.+1640d-2652)t\right]\sunsetd(s)
\nonumber \\ &&
+3\frac{(d-3)(3d-8)(3d-10)}{(d-4)^3(d-5)^2(d-6)s^2t^2}
\nonumber \\ &&
\times\left[(16d^3-229d^2+1088d-1716)s\right.
\nonumber \\ &&
\left.+(d-5)(3d-14)(3d-16)t\right]\sunsetd(t) ,
\end{eqnarray}
where
\begin{eqnarray}
\label{eq:pinchdef}
\diagb^d(s, t) &=& I \left( 0,\, 1,\, 1,\, 0,\, 1,\, 1,\, 1,\, 0;\, d\right),
\nonumber \\
\blobb^d(s, t) &=& I \left( 1,\, 1,\, 1,\, 0,\, 1,\, 0,\, 1,\, 0;\, d\right),
\nonumber \\
\spec^d(s)     &=& I \left( 1,\, 0,\, 1,\, 1,\, 0,\, 1,\, 0,\, 0;\, d\right),
\nonumber \\
\suns^d(s)     &=& I \left( 0,\, 0,\, 1,\, 0,\, 0,\, 1,\, 1,\, 0;\, d\right),
\nonumber \\
\suns^d(t)     &=& I \left( 0,\, 1,\, 0,\, 0,\, 1,\, 0,\, 1,\, 0;\, d\right),
\nonumber \\
\tr^d(s)       &=& I \left( 1,\, 0,\, 1,\, 0,\, 1,\, 0,\, 1,\, 0;\, d\right).
\end{eqnarray}
The factors of $(d-4)$ in the denominators of the first two terms on
the right hand side of eq.~(\ref{eq:relmasters}) are the ones we are
concerned about.

The purpose of this paper is to calculate the integral with
the irreducible numerator $\nigel$ directly from a Mellin-Barnes
representation. It can then be used instead of $\veretin$ as a new master
integral. We check our result in two different ways: firstly, by verifying
that $\smirnov$ and $\nigel$ satisfy a system of differential equations,
and secondly, by using them to compute $\smirnov$ and $\veretin$ in $d=6$
dimensions, both of which are finite, and comparing the result with a
numerical integration.
\markright{}

\section{Calculation by Mellin-Barnes contour integrals}

The analytic structure of the on-shell double box is rather simple, since
it only depends on two scales, and its only thresholds are at $s=0$ and
$t=0$. The main difficulty in calculating this diagram is to find a way
to isolate its infrared and collinear divergences. A general recursive
solution to this problem based on Feynman parameters was proposed in
ref.~\cite{HeinrichBinoth}. However, for the analytical calculation we
are going to present here, it is more convenient to use a Mellin-Barnes
representation~\cite{Smirnovdoublebox,Ussyukina}, which enables us to
isolate the poles in $\epsilon$ in a very natural way.

We derive our Mellin-Barnes representation for the two-loop
integrals (\ref{eq:Idef}) by doing the loop integrations
one by one (cf. the off-shell calculation~\cite{UDdoublebox}).
In terms of Feynman parameters, the $l$-loop
can be written as
\begin{eqnarray}
\label{eq:l-loopFP}
\lefteqn{
\int \mbox{d}^d l \;
\frac{1}{P_4^{\nu_4} P_5^{\nu_5} P_6^{\nu_6} P_7^{\nu_7}} =
 i \pi^{d/2} (-1)^{\nu_{4567}} 
}
&& \nonumber \\ && \times
\frac{ \Gamma( \nu_{4567}-d/2 ) }
{ \Gamma(\nu_4) \Gamma(\nu_5) \Gamma(\nu_6) \Gamma(\nu_7) }
 \int \limits_{0}^{\infty}
\prod \limits_{j=4}^7 \mbox{d} x_j \, x_j^{\nu_j-1}
\nonumber \\ && \times
 \delta(1-x_4-x_5-x_6-x_7) \,  C^{d/2-\nu_{4567}} ,
\end{eqnarray}
where
$\nu_{4567}=\nu_4+\nu_5+\nu_6+\nu_7$ (similar
abbreviations will be used below), and
\begin{equation}
C = -P_1 x_6 x_7 - P_3 x_4 x_7 -P_8 x_5 x_7 - s x_4 x_6 .
\end{equation}
By introducing three Mellin-Barnes parameters,
$\alpha$, $\beta$ and $\tau$, we split the polynomial
$C$ into factors:
\begin{eqnarray}
\label{eq:C-MB}
\lefteqn{
\Gamma( \nu_{4567}-d/2 ) \,C^{d/2-\nu_{4567}} =
 \int \limits_{-i\infty}^{i\infty} \frac{ \mbox{d} \alpha}{2 \pi i}
 \int \limits_{-i\infty}^{i\infty} \frac{ \mbox{d} \beta }{2 \pi i}
}
&& \nonumber \\ &&
 \int \limits_{-i\infty}^{i\infty} \frac{ \mbox{d} \tau  }{2 \pi i}
 (-P_1 x_6 x_7)^{\alpha}
 (-P_3 x_4 x_7)^{\beta}
 (-P_8 x_5 x_7)^{\tau}
\nonumber \\ && \times
 (-s x_4 x_6)^{d/2-\nu_{4567}-\alpha-\beta-\tau}
\Gamma(-\alpha) \Gamma(-\beta)
\nonumber \\ && \times
\Gamma(-\tau)
\Gamma(\nu_{4567}-d/2+\alpha+\beta+\tau) .
\end{eqnarray}
After inserting eq.~(\ref{eq:C-MB}) into eq.~(\ref{eq:l-loopFP}),
we evaluate the Feynman parameter integrals in terms of
gamma functions, which gives us the following Mellin-Barnes
representation for the $l$-loop:
\begin{eqnarray}
\label{eq:l-loop}
\lefteqn{
\int \mbox{d}^d l \;
\frac{1}{P_4^{\nu_4} P_5^{\nu_5} P_6^{\nu_6} P_7^{\nu_7}} =
\frac{ i \pi^{d/2} (-1)^{\nu_{4567}} }
{ \Gamma(\nu_4) \Gamma(\nu_5) \Gamma(\nu_6) \Gamma(\nu_7) }
} && \nonumber \\ \lefteqn{ \times
\frac{1}{\Gamma(d-\nu_{4567})}
 \int \limits_{-i\infty}^{i\infty} \frac{ \mbox{d} \alpha}{2 \pi i}
 \int \limits_{-i\infty}^{i\infty} \frac{ \mbox{d} \beta }{2 \pi i}
 \int \limits_{-i\infty}^{i\infty} \frac{ \mbox{d} \tau  }{2 \pi i}
} && \nonumber \\ \lefteqn{ \times
 (-P_1)^{\alpha}
 (-P_3)^{\beta}
 (-P_8)^{\tau}
 (-s)^{d/2-\nu_{4567}-\alpha-\beta-\tau}
} && \nonumber \\ \lefteqn{ \times
\Gamma(-\alpha) \Gamma(-\beta) \Gamma(-\tau)
\Gamma(\nu_{4567}-d/2+\alpha+\beta+\tau)
} && \nonumber \\ \lefteqn{ \times
\Gamma(d/2-\nu_{567}-\alpha-\tau)
\Gamma(d/2-\nu_{457}-\beta-\tau)
} && \nonumber \\ \lefteqn{ \times
\Gamma(\nu_5+\tau) \Gamma(\nu_7+\alpha+\beta+\tau) .
} &&
\end{eqnarray}
When this result is inserted into (\ref{eq:Idef}), the remaining
$k$-integral has the form of an on-shell massless one-loop box diagram
with indices $\nu_1-\alpha$, $\nu_2$, $\nu_3-\beta$, $\nu_8-\tau$.
We repeat the above steps for this $k$-integral, using
a further Mellin-Barnes parameter, $\sigma$, and finally
obtain
\begin{eqnarray}
\label{eq:MB-general}
\lefteqn{
I(\nu_1,\nu_2,\nu_3,\nu_4,\nu_5,\nu_6,\nu_7,\nu_8;d) =
} && \nonumber \\ \lefteqn{
\frac{ {(i \pi^{d/2})}^2 (-1)^{\nu_{12345678} } }
{ \Gamma(\nu_2) \Gamma(\nu_4) \Gamma(\nu_5) \Gamma(\nu_6) \Gamma(\nu_7)
\Gamma(d-\nu_{4567}) }
\frac{1}{(2 \pi i)^4}
} && \nonumber \\ \lefteqn{ \times
 \int \limits_{-i\infty}^{i\infty} \mbox{d} \alpha
 \int \limits_{-i\infty}^{i\infty} \mbox{d} \beta
 \int \limits_{-i\infty}^{i\infty} \mbox{d} \tau
 \int \limits_{-i\infty}^{i\infty} \mbox{d} \sigma
(-t)^{-\sigma}
(-s)^{d-\nu+\sigma}
} && \nonumber \\ \lefteqn{ \times
\frac{ 
\Gamma(-\alpha) \Gamma(-\beta) \Gamma(-\tau)
\Gamma(\nu_{4567}-d/2+\alpha+\beta+\tau) }
{ \Gamma(\nu_1-\alpha) \Gamma(\nu_3-\beta) \Gamma(\nu_8-\tau)}
} && \nonumber \\ \lefteqn{ \times
\frac{\Gamma(\sigma) \Gamma(\nu_{1238}-d/2-\alpha-\beta-\tau-\sigma)}
{ \Gamma(d-\nu_{1238}+\alpha+\beta+\tau) }
} && \nonumber \\ \lefteqn{ \times
\Gamma(d/2-\nu_{567}-\alpha-\tau)
\Gamma(d/2-\nu_{457}-\beta-\tau)
} && \nonumber \\ \lefteqn{ \times
\Gamma(\nu_5+\tau) \Gamma(\nu_7+\alpha+\beta+\tau)
} && \nonumber \\ \lefteqn{ \times
\Gamma(d/2-\nu_{128}+\alpha+\tau+\sigma)
\Gamma(\nu_8-\tau-\sigma)
} && \nonumber \\ \lefteqn{ \times
\Gamma(d/2-\nu_{238}+\beta+\tau+\sigma)
\Gamma(\nu_2-\sigma) .
} &&
\end{eqnarray}
In deriving this formula, we have assumed that the various parameters
are such, that all the manipulations we performed are justified. This
is certainly true if we are able to find a set of straight lines,
parallel to the imaginary axis, for the integration variables $\alpha$,
$\beta$, $\sigma$, and $\tau$, such that the arguments of all the
gamma functions in it have positive real parts. We then define the
integrals~(\ref{eq:Idef}) for values of the parameters where such contours
do not exist by analytic continuation.

Let us now consider the case with the irreducible numerator,
$\nigel$. On the one hand, from the definition~(\ref{eq:I3def}),
$\nu_5+\nu_8=1-1=0$. On the other hand, if the real parts of the
arguments of all gamma functions are positive, then in particular
$Re(\nu_5+\tau)$, $Re(\sigma)$ and $Re(\nu_8-\tau-\sigma)$ are positive,
and therefore $Re(\nu_5+\nu_8) > 0$. Since this does not depend on $d$,
it means that in order to calculate $\nigel$ using the Mellin-Barnes
representation~(\ref{eq:MB-general}), we must perform an analytic
continuation not only in $d$, but also in $\nu_5$ or $\nu_8$. We choose
$\nu_8$. Setting $\nu_8=-1+\eta$ and all other $\nu$'s equal to one,
we get
\begin{eqnarray}
\label{eq:MB-I3}
\lefteqn{ \nigel = \lim_{\eta \downarrow 0}
I(1,1,1,1,1,1,-1+\eta; d) =
\frac{ {(i \pi^{d/2})}^2 }{\Gamma(-2\epsilon)}
} && \nonumber \\ \lefteqn{ \times
 \lim_{\eta \downarrow 0}\frac{1}{(2 \pi i)^4}
 \int \limits_{-i\infty}^{i\infty} \mbox{d} \alpha
 \int \limits_{-i\infty}^{i\infty} \mbox{d} \beta
 \int \limits_{-i\infty}^{i\infty} \mbox{d} \tau
 \int \limits_{-i\infty}^{i\infty} \mbox{d} \sigma
(-t)^{-\sigma}
} && \nonumber \\ \lefteqn{ \times
(-s)^{-2-\eta-2\epsilon+\sigma}
\Gamma(\sigma) \Gamma(1-\sigma)
} && \nonumber \\ \lefteqn{ \times
\frac{ 
\Gamma(-\alpha) \Gamma(-\beta) \Gamma(-\tau) \Gamma(1+\tau)
\Gamma(-1+\eta-\tau-\sigma)
 }
{ \Gamma(1-\alpha) \Gamma(1-\beta) \Gamma(-1+\eta-\tau)}
} && \nonumber \\ \lefteqn{ \times
\frac{
\Gamma(1+\alpha+\beta+\tau)
\Gamma(2+\epsilon+\alpha+\beta+\tau)
}
{ \Gamma(2-\eta-2\epsilon+\alpha+\beta+\tau) }
} && \nonumber \\ \lefteqn{ \times
\Gamma(-1-\epsilon-\alpha-\tau)
\Gamma(1-\eta-\epsilon+\alpha+\tau+\sigma)
} && \nonumber \\ \lefteqn{ \times
\Gamma(-1-\epsilon-\beta-\tau)
\Gamma(1-\eta-\epsilon+\beta+\tau+\sigma)
} && \nonumber \\ \lefteqn{ \times
\Gamma(\eta+\epsilon-\alpha-\beta-\tau-\sigma) .
} &&
\end{eqnarray}
We can make the real parts of the arguments of all gamma functions in
eq.~(\ref{eq:MB-I3}) positive by picking, for example, $\eta=12 y$ and
$\epsilon=-12 y$, where $y$ is some positive number much smaller than
one, and choosing contours for the Mellin-Barnes variables defined by:
$Re(\alpha)=Re(\beta)=-y$, $Re(\tau)=-1+4 y$ and $Re(\sigma)=4 y$.

Starting from these values, we first perform an analytic continuation
in $\eta$ to $\eta=0$, keeping $\epsilon$ fixed, and then another one
in $\epsilon$ to the vicinity of $\epsilon=0$. The procedure for both
continuations is straightforward: keeping the integration contours fixed,
we simply have to keep track of the poles of the gamma functions, and
whenever one of them crosses an integration contour, add its residue to a
list of terms that will contribute to the final answer. For example, with
the above choice of contours, the first crossing happens at $\eta=8y$,
when the pole at $\tau=-1+\eta-\sigma$ crosses the $\tau$-contour.
After taking a residue in one integration variable, we continue to follow
the poles in the remaining variables, building up a tree of single and
multiple residue terms. By this procedure, poles in $\epsilon$ are
automatically expressed through singular gamma functions multiplying
integrals that can safely be expanded under the integral sign.

To ${\cal O}(\epsilon^0)$, it turns out that, along with terms where there
is no integral left, only single and two-fold integrals contribute, because
terms with more integrals are killed by the factor $1/\Gamma(-2\epsilon)$
in eq.~(\ref{eq:MB-I3}). In the two-fold integrals, one integration
can be done by Barnes's first lemma~\cite{Slater}. The single integrals
that are left can all be calculated by closing the contour and summing
harmonic series. We find the following result:
\begin{eqnarray}
\label{eq:I3-res}
\lefteqn{
\nigel =
\frac{ (i \pi^{d/2})^2 \Gamma(1+\epsilon)^2
} { s^2 (-s)^{2 \epsilon} }
\left\{ \frac{9}{4\,\epsilon^4} - \frac{2}{\epsilon^3} \ell
-\frac{7 \pi^2}{3\,\epsilon^2}
\right.
} && \nonumber \\ \lefteqn{
{}+\frac{1}{\epsilon} \left[
\frac{4}{3} \ell^3 + \frac{14}{3} \pi^2 \ell
-4 \left( \ell^2 + \pi^2 \right) \log(1+t/s)
\right.
} && \nonumber \\ \lefteqn{
\hspace{2 em}
{} + 8 \,\Li{3}{-t/s}
-8 \,\ell \,\Li{2}{-t/s}
-16 \,\zeta(3)
\left. \vphantom{\frac{1}{1}} \right]
} && \nonumber \\ \lefteqn{
{}-\frac{4}{3} \ell^4
-\frac{13}{3} \pi^2 \ell^2
+\left( \frac{16}{3} \ell^3 +\frac{26}{3} \pi^2 \ell \right)\log(1+t/s)
} && \nonumber \\ \lefteqn{
{}-5 \left( \ell^2 + \pi^2 \right) \log^2(1+t/s)
} && \nonumber \\ \lefteqn{
{}+ \left( 6\,\ell^2 - 20\,\ell \log(1+t/s) - \frac{4}{3} \pi^2 \right)
\Li{2}{-t/s}
} && \nonumber \\ \lefteqn{
{}+ \left( 8\,\ell + 20 \log(1+t/s) \right) \Li{3}{-t/s}
} && \nonumber \\ \lefteqn{
{}+ 20\,\snp{2,2}{-t/s}
- 20 \,\ell \,\snp{1,2}{-t/s}
-28 \,\Li{4}{-t/s}
} && \nonumber \\ \lefteqn{
\left.
{}+ \left( 28 \,\ell - 20 \log(1+t/s) \right) \zeta(3)
- \frac{7 \pi^4}{45}
\right\} ,
} &&
\end{eqnarray}
with $\ell=\log(t/s)$.

\section{Differential equations for double-box master integrals}

The integrals $\smirnov$ and $\nigel$ can now be chosen as a new basis of
master integrals of the planar double-box topology, solving the problem
of the ${\cal O} (1/\epsilon)$ coefficients of $\smirnov$ and $\veretin$
in matrix element calculations~\cite{nigel}.

It is possible to construct a system of differential equations satisfied by
the double box master integrals \cite{SmirnovVeretin99,GehrmannRemiddi}.
In terms of the new basis, these differential equations are \cite{nigel}:
\begin{eqnarray}
\label{eq:ddtsmirnov}
\lefteqn{
\frac{\partial}{\partial t}\svboxad(s,t) =
\frac{[(d-5)s-t]}{(s+t)t}\svboxad(s,t)
}
\nonumber \\ &&
+\frac{(d-4)}{(s+t)t}\nigelboxd(s,t)
\nonumber \\ &&
-6\frac{(d-4)}{st^2}\diagboxd(s,t)
\nonumber \\ &&
+12\frac{(d-3)}{(s+t)t^2}\bubboxd(s,t)
\nonumber \\ &&
+4\frac{(d-3)^2}{(d-4)s^2t(s+t)}\glassd(s)
\nonumber \\ &&
+3\frac{(d-3)(3d-10)(2s+t)}{(d-4)s^2t^2(s+t)}\trid(s)
\nonumber \\ &&
+6\frac{(d-3)(3d-8)(3d-10)(s-t)}{(d-4)^2s^3t^2(s+t)}\sunsetd(s)
\nonumber \\ &&
+6\frac{(d-3)(3d-8)(3d-10)}{(d-4)^2st^3(s+t)}\sunsetd(t)
\end{eqnarray}
\begin{eqnarray}
\label{eq:ddtnigel}
\lefteqn{
\frac{\partial}{\partial t}\nigelboxd(s,t) =
-\frac{1}{2}\frac{(d-4)s}{(s+t)t}\nigelboxd(s,t)
}
\nonumber \\ &&
+\frac{1}{2}\frac{(d-4)s}{s+t}\svboxad(s,t)
\nonumber \\ &&
-9\frac{(d-4)}{st}\diagboxd(s,t)
\nonumber \\ &&
+12\frac{(d-3)}{(s+t)t}\bubboxd(s,t)
\nonumber \\ &&
+2\frac{(d-3)^2(s+2t)}{(d-4)s^2t(s+t)}\glassd(s)
\nonumber \\ &&
+\frac{15}{2}\frac{(d-3)(3d-10)}{(d-4)st(s+t)}\trid(s)
\nonumber \\ &&
+6\frac{(d-3)(3d-8)(3d-10)}{(d-4)^2s^2t(s+t)}\sunsetd(s)
\nonumber \\ &&
+9\frac{(d-3)(3d-8)(3d-10)}{(d-4)^2st^2(s+t)}\sunsetd(t)
\end{eqnarray}
Expanding eqs.~(\ref{eq:ddtsmirnov},\ref{eq:ddtnigel}) in $\epsilon$, and
inserting the expansion of $\smirnov$ from ref.~\cite{Smirnovdoublebox},
of the pinched diagrams~(\ref{eq:pinchdef}) (see, eg. \cite{babisetal}),
and the result of eq.~(\ref{eq:I3-res}), we find that they are indeed
satisfied.

Inspecting the right hand sides of the differential equations,
one notices that in eq.~(\ref{eq:ddtsmirnov}), the coefficient
of $\nigel$, and in eq.~(\ref{eq:ddtnigel}), those of $\smirnov$ and
$\nigel$, are all proportional to $d-4$. This means that, if $\smirnov$
is known to ${\cal O}(\epsilon^0)$, the ${\cal O}(\epsilon^0)$ part
of $\nigel$ is, a priori, only determined by the system of equations up
to a $t$-independent constant.

One also observes that the system of differential equations has
a singular point at $s+t=0$. This corresponds to the special
kinematical configuration where $p_1+p_3=0$. At this point,
the numerator of $\nigel$ becomes reducible:
\begin{eqnarray}
P_8 & = & (k+p_1+p_2+p_3)^2
\nonumber \\ & = & (k+p_2)^2 = P_1-P_2+P_3-s ,
\end{eqnarray}
so $\nigel$ collapses to a linear combination of $\smirnov$ and pinched
diagrams.

For a further discussion of this singular point, we refer to the
contribution of Gehrmann and Remiddi~\cite{GehrmannRemiddiLL}, who
recently investigated the (equivalent) system of differential equations
for $\smirnov$ and $\veretin$. By excluding singular solutions of
the differential equations as possible candidates, they were able to
extract the order $\epsilon$ piece of $\smirnov-t\veretin$ using only
the value of $\smirnov$ at $s+t=0$ to ${\cal O}(\epsilon^0)$
as input. The information contained in this quantity is equivalent to
our eq.~(\ref{eq:I3-res}).

\section{6 dimensions}

The master integrals $\smirnov$ and $\veretin$ are both finite in
$d=6$ dimensions \cite{BernDDPRozowsky98}. This can be deduced 
from power counting considerations in momentum space; it is
also easy to see by examining the arguments of the gamma functions
in the Mellin-Barnes representation~(\ref{eq:MB-general}).
Following the method of \cite{SmirnovVeretin99}, we relate
these master integrals in $d=6-2 \epsilon$ dimensions to
master integrals of the new basis in $d=4-2 \epsilon$ dimensions.
Substituting $\epsilon$-expansions for the latter, we find that all
pole terms indeed cancel, and the finite parts are
\begin{eqnarray}
\label{eq:smirnovbox6d}
\smirnov^{d=6} & = & -\pi^6 \left\{
\frac{a_1}{(s+t)} + \frac{b}{t} \right\} ,
\\
\label{eq:veretinbox6d}
\veretin^{d=6} & = & -\pi^6 \left\{
\frac{a_2}{s (s+t)} + \frac{b}{s\,t} \right\} ,
\end{eqnarray}
\begin{eqnarray}
a_1 & = &
 t \, \frac{\partial a_2}{\partial t} - 6\,\zeta(3) ,
\end{eqnarray}
\begin{eqnarray}
a_2 \!\! & = & \!\!
  \frac{\pi^4}{10}
    + 6\,\Li{4}{-t/s}
    + \left( \ell^2+\pi^2 \right)\,\Li{2}{-t/s}
\nonumber \\ &&
{}  + 4\,\ell \left(\zeta(3)-\Li{3}{-t/s} \right)
    + \frac{\pi^2}{6} \ell^2 ,
\end{eqnarray}
\begin{eqnarray}
b \!\! & = & \!\!
\left( 2 \,\zeta(3)
     - 2 \,\Li{3}{-t/s}
     - \frac{\pi^2}{3} \ell \right) \log(1+t/s) 
\nonumber \\ && 
{} + \frac{1}{2} ( \ell^2 + \pi^2) \log^2(1+t/s)
\nonumber \\ && 
{} + \left( 2 \,\ell\,\log(1+t/s)
   - \frac{\pi^2}{3} \right) \,\Li{2}{-t/s}
\nonumber \\ && 
{} + 2 \,\ell \,\snp{1,2}{-t/s}
 - 2 \,\snp{2,2}{-t/s} .
\end{eqnarray}
We checked these expressions by numerical integration of
Feynman parametric representations for the integrals in $d=6$
in the region of negative $s$ and $t$. This gives us an
independent check on the finite parts of our result
for $\nigel$, eq.~(\ref{eq:I3-res}), and of $\smirnov$
in ref.~\cite{Smirnovdoublebox}.

\section{Conclusion}
We have resolved a technical issue related to the reduction of on-shell
massless double box diagrams to master integrals.  Looking back,
one can say that the basis of master integrals
$\{\smirnov, \, \veretin\}$ used in ref.~\cite{SmirnovVeretin99}
was an unfortunate choice for applications in $d=4-2\epsilon$
dimensions, because it leads to factors of $1/\epsilon$ in the
reduction coefficients. The troublesome terms do not cancel in physical
amplitudes \cite{nigel}. If one uses, instead, the basis $\{\smirnov,
\, \nigel\}$, the factors of $1/\epsilon$ disappear. Our main result
is the expansion of the integral $\nigel$ in terms of polylogarithms
to ${\cal O}(\epsilon^0)$, eq.~(\ref{eq:I3-res}), which we derived from
scratch using a Mellin-Barnes representation.  (An equivalent result was
obtained independently by Gehrmann and Remiddi \cite{GehrmannRemiddiLL},
by solving a system of differential equations and using the first master
integral from ref.~\cite{Smirnovdoublebox} as input). We verified
that our result, together with the first master integral, satisfies
the differential equations~(\ref{eq:ddtsmirnov},\ref{eq:ddtnigel}).
As a further check on the constant which is not determined by the
differential equations alone, we used our result to derive analytical
expressions for two finite double box integrals in $d=6$ dimensions,
eqs.~(\ref{eq:smirnovbox6d},\ref{eq:veretinbox6d}), and checked them by
numerical integration.

\subsection*{Acknowledgements}
We thank E.W.N.~Glover, D.J.~Broadhurst and O.L.~Veretin for discussions
and useful suggestions. C.A. acknowledges financial support from
the Greek Government, J.B.T. acknowledges financial support from
the DFG-Forschergruppe ``Quantenfeldtheorie, Computeralgebra und
Monte-Carlo-Simulation'', and M.E.T. acknowledges financial support
from CONACyT and the CVCP. We gratefully acknowledge the support of
the British Council and German Academic Exchange Service under ARC
project 1050.


\end{document}